\documentclass[IOP]{emulateapj}
\usepackage{amssymb}
\usepackage{longtable}
\usepackage{amsmath}
\newcommand{\DM}{{\rm DM}}
\newcommand{\code}[1]{{\tt #1}}

\shorttitle{Dark Matter Distribution in Abell 383}
\submitted{Accepted to ApJL}
\begin{document}

\shortauthors{Newman et al.}

\title{The Dark Matter Distribution in Abell 383: Evidence for a Shallow Density Cusp
from Improved Lensing, Stellar Kinematic and X-ray Data}

\author {Andrew B. Newman\altaffilmark{1},  Tommaso Treu\altaffilmark{2},
Richard S. Ellis\altaffilmark{1}, David J. Sand\altaffilmark{2,3} }

 \altaffiltext{1}{Cahill Center for Astronomy \& Astrophysics, California
  Institute of Technology, MS 249-17, Pasadena, CA 91125}
  \altaffiltext{2}{Department of Physics, University of California, Santa Barbara, CA 93106}
\altaffiltext{3}{Las Cumbres Observatory Global Telescope Network, Santa Barbara, CA 93117}

\begin{abstract}
 
We extend our analyses of the dark matter (DM) distribution in relaxed clusters to the case of Abell
383, a luminous X-ray cluster at $z$=0.189 with a dominant central galaxy and numerous
strongly-lensed features. Following our earlier papers, we combine strong and weak lensing constraints
secured with \emph{Hubble Space Telescope} and \emph{Subaru} imaging
with the radial profile of the stellar velocity dispersion of the central galaxy, essential for separating
the baryonic mass distribution in the cluster core. Hydrostatic mass estimates from \emph{Chandra} X-ray observations 
further constrain the solution. These combined datasets provide nearly continuous constraints extending from
2~kpc to 1.5~Mpc in radius, allowing stringent tests of results from recent numerical simulations.
Two key improvements in our data and its analysis make this the most robust case yet for a shallow slope
$\beta$ of the DM density profile $\rho_{\DM}\propto r^{-\beta}$ on small scales.
First, following deep Keck spectroscopy, we have secured the stellar velocity dispersion profile
to a radius of 26~kpc for the first time in a lensing cluster.
Secondly, we improve our previous analysis by adopting a triaxial DM distribution and axisymmetric dynamical models.
We demonstate that in this remarkably well-constrained system, the logarithmic slope of the DM density
at small radii is $\beta < 1.0$ (95\% confidence).
An improved treatment of baryonic physics is necessary, but possibly insufficient, to reconcile
our observations with the recent results of high-resolution simulations.

\end{abstract} 

\section{Introduction}
\label{sec:intro}

The cold dark matter (CDM) paradigm has been remarkably successful at predicting
the large-scale distribution of matter in the Universe as well as its observed evolution
from the earliest epochs to the present day \citep{Springel06}. However, much interest
has been shown in possible discrepancies that remain on small scales between its predictions and 
the available observations. A source of continuing puzzlement relates to the observed form 
of the dark matter (DM) profile seen in galaxy clusters. 

Numerical simulations predict CDM halos follow a self-similar
density profile whose three-dimensional (3D) form within a scale radius $r_s$ approaches a 
cusp $\rho \propto r^{-\beta}$ with an inner slope $\beta \simeq 1 - 1.3$ at the smallest
resolved scales \citep[e.g.,][]{NFW96,Ghigna00,Diemand05}.
Improved resolution has suggested modest flattening \citep{Navarro10}, but only at very small radii.
This is in contradiction to observations based on the combination of 
strong lensing and stellar kinematics which yielded much flatter inner slopes of $\beta \simeq 0.5$ 
for two well-studied clusters \citep{Sand08}. On the other hand, a steeper $\beta = 0.92 \pm 0.04$ has been inferred
in Abell~1703 \citep{Richard09}, possibly indicating significant
scatter in the inner structure of clusters. 

Recently in \citet[][hereafter N09]{Newman09} we further developed the method introduced
by \citet{Sand04,Sand08} by incorporating weak lensing constraints on the large-scale mass distribution
using \emph{Subaru} imaging of Abell 611. This removed
a degeneracy between the scale radius $r_s$ and $\beta$ and led to the first measurement
of the DM profile across a dynamic range in cluster-centric radius (3~kpc to 3.3~Mpc) comparable to that 
presently achieved in simulations. A shallow cusp with $\beta < 0.3$ (68\%) was derived.

Here we make two further improvements
in our methodology and apply these to Abell 383 ($z=0.189$). 
This cluster has a regular optical and X-ray morphology, a remarkably low sub-structure
fraction, and a dominant, near-circular brightest cluster galaxy (BCG) \citep{Smith05,Smith08}.
\citet{Sand04} initially studied this cluster assuming spherical symmetry but later
\citep{Sand08} undertook a two-dimensional (2D) lensing analysis, deriving $\beta = 0.45 \pm 0.2$. In
addition to removing the scale radius degeneracy discussed above, we have 
considerably extended the range of the stellar kinematic constraints via a deep
Keck spectrum of the BCG, significantly improving our knowledge of the mass distribution
on $\lesssim 30$~kpc scales.  Secondly, we use
\emph{Chandra} X-ray data to determine the line-of-sight ellipticity in the mass distribution,
thereby achieving constraints on a 3D model with minimal
uncertainties arising from projection effects. 

We adopt a cosmology
with  $\Omega_{\Lambda}=0.7$, $\Omega_{\rm m}=0.3$ and
$H_{0}=70\,h_{70}~{\rm km\,s}^{-1}\,{\rm Mpc}^{-1}$ throughout.

\section{Observational Ingredients}

We first discuss in turn the four observational ingredients we use to constrain
the distribution of dark matter and baryons in Abell 383. 

\begin{figure}
\plotone{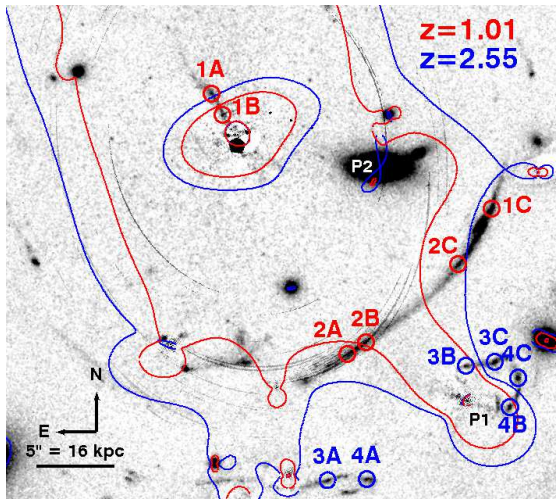}
\caption{Strong lensing constraints: \emph{HST}/WFC2 image in the F702W filter
with the BCG and other cluster galaxies subtracted for clarity. Two families of multiple
image systems with known spectroscopic redshifts, each comprising two sources, are
identified in the legend.
Critical lines are plotted for both source planes.}
\label{fig:strong}
\end{figure}

\subsection{Strong Lensing}

Figure~\ref{fig:strong} shows the multiply-imaged sources, tabulated
in \citet{Sand04,Sand08} and \citet{Smith05},
which comprise the strong lensing constraints:
a radial/tangential arc at $z_{\rm spec}=1.01$ and
a complex system in the southwest. In previous analyses the redshift of the
latter system was unknown, but following Keck/LRIS observations in October~2009
we secured a spectroscopic redshift $z_{\rm spec}=2.55$ for images 3C and 4C.
The $n_i$=12 images of $n_s$=4 sources produces $2(n_i - n_s) = 16$
constraints. Following N09, we use
\code{Lenstool}\footnote{\url{http://www.oamp.fr/cosmology/lenstool/}} \citep{Jullo07}
in source plane mode for the strong lensing analysis.
Consistent with earlier work, we adopt an uncertainty of
$\sigma_{\rm pos} = 0\farcs5$ for the image positions to account for systematic modelling
uncertainty.

\subsection{Weak Lensing}

The large-scale shear arising from Abell 383 has been analyzed with
multi-color imaging taken using SuprimeCam at the {\it Subaru} telescope. The
shear was measured in $R_C$ images taken by the authors on
12-13 November 2007 in excellent seeing of $0\farcs57$.
Broad-band photometry from $BViz$ images in the {\it Subaru} archive was used with the
\code{BPZ} code \citep{BPZ} to obtain photometric redshifts.
The procedures closely followed those discussed by N09. 
From a sample of galaxies with $5\sigma$ detections in $R_C$, a population with $0.5 <z_b< 2$
was selected for shear measurement, yielding a surface density of 25~arcmin$^{-2}$.
As discussed in N09,
our shear measurements were calibrated using the
recovery factor $m_{\rm WL} = 0.81 \pm 0.04$ based on the STEP2 simulations \citep{Massey07}.

\begin{figure*}
\epsscale{0.7}
\plotone{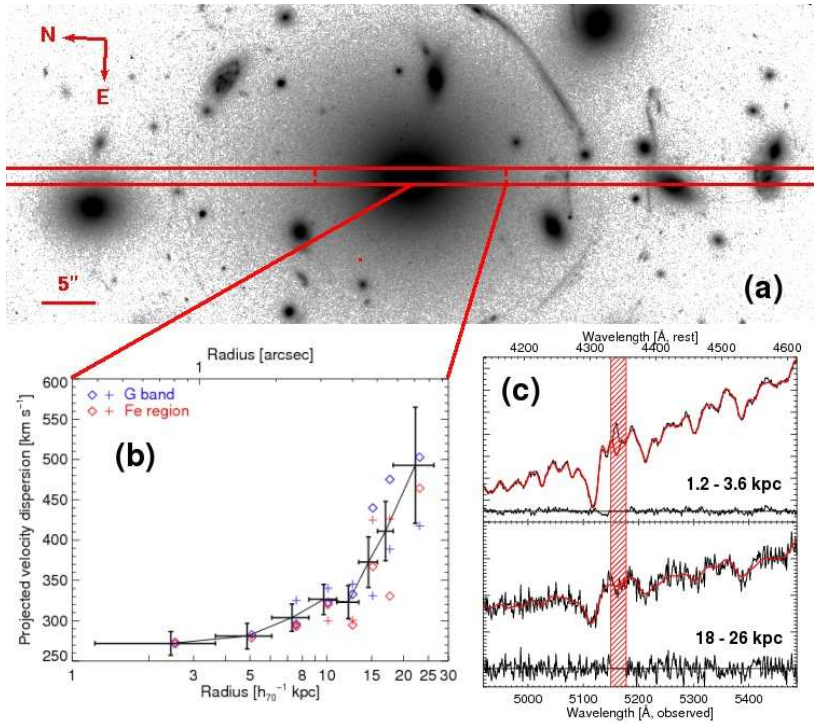}
\caption{Considerably improved kinematic data for the central galaxy. (a) 
Long slit configuration for the 22.8~ks LRIS exposure; the $8\farcs8$ radial extent
over which velocity dispersions could be derived is indicated by the vertical 
red line. (b)
Averaged stellar velocity dispersion profile as derived independently
from the G band and Fe $\lambda5270$ absorption line regions (blue and red symbols) on either side of the center (diamond
and plus symbols).
The rapid rise in the velocity dispersion at large radii due to the DM halo is clearly evident.
(c) Spectra for the inner- and outermost spatial bins around the G band. Red curves show the fits to the
broadened stellar template, with residuals plotted below.}
\label{fig:dynamics}
\end{figure*}

\subsection{Stellar kinematics}

We have substantially improved the data used by \citet{Sand08} by securing a much deeper spectroscopic exposure of the BCG
(Figure~\ref{fig:dynamics}). Earlier data 
yielded a stellar velocity dispersion profile $\sigma_{los}(R)$ extending to $R=5$~kpc in 3 spatial bins
(Figure~4 of \citealt{Sand08}). The present data comprise a 22.8~ks integration with Keck/LRIS taken on
12 October 2009 using the 600 mm${}^{-1}$ blue grism and the 600 mm${}^{-1}$ red grating
blazed at 7500~\AA.
The $1\farcs5$ slit yielded a resolution of $\sigma_{\rm inst} = 153$~ km~s$^{-1}$ at the G band.

The improved depth of this exposure has enabled us to secure a reliable dispersion profile to 26 kpc (circularized), which
can be verified independently using both the G band and Fe absorption lines (Figure~\ref{fig:dynamics}).
Spectra of G and K giants from the MILES library \citep{SanchezBlazquez06}
were used to synthesize the optimal stellar template \citep[see][]{Newman10}.
The gain over earlier data is substantial both in extent and sampling. Although $\sigma_{los}(R)$ has been
measured to very large radii in local cD galaxies \citep{Kelson02}, this is to our knowledge the most 
extended measurement yet obtained in a lensing cluster.

\subsection{X-ray}
\label{sec:Xray}

The final ingredient is the mass distribution probed by X-ray emission from the intracluster medium.
This was measured by \citet[][hereafter A08]{Allen08} using {\it Chandra} data. Although their analysis
assumed spherical symmetry, this has very little effect on the inferred \emph{spherically-averaged} mass profile,
as we discuss further in Section~\ref{sec:systematics}. Mock observations of simulated clusters show that non-thermal
sources of pressure cause X-ray--derived masses to be biased by $\simeq -10\%$ \citep{Nagai07,Lau09}. To account for
this, we place a Gaussian prior on $m_{\rm X} \equiv M_{\rm HSE} / M_{\rm true} = 0.9 \pm 0.1$, where $M_{\rm HSE}$ represents
the A08 results and $M_{\rm true}$ the true spherically-averaged mass distribution. The 10\% uncertainty in $m_{\rm X}$ reflects
the cluster-to-cluster scatter in non-thermal support, as well as uncertainty in the {\it Chandra} temperature
calibration \citep{Reese10}. From the A08 mass profile, we take five points spaced log-uniformly
from $50-600$~kpc to match the number of independent temperature measurements. (The results are not sensitive to the
inner limit.) Finally, we add 10\% in quadrature to the uncertainty in each data point to allow for systematic
errors with radial gradients (e.g., non-thermal pressure and errors arising from non-sphericity). 

\section{Deriving the Mass Distribution}
\label{sec:methods}

Our model of the cluster mass distribution comprises three components: 
(1) the cluster-scale halo, (2) stellar mass in the BCG, and (3) mass in other cluster
galaxies which are incorporated as perturbations in the lensing
analysis.  The third component is modelled as described in our
previous work, including two individually-modelled galaxies
(P1 and P2 in Figure~\ref{fig:strong}).
Following N09, the cluster halo and BCG
are described by generalized NFW (gNFW) and dual
pseudoisothermal ellipsoid (dPIE) profiles, with a key
improvement: the newly-incorporated X-ray data
allows us to consider triaxial mass models.

By combining X-ray and lensing constraints, we can directly measure the elongation of the
DM distribution along the line of sight (l.o.s), thus addressing a key
systematic uncertainty in deriving the mass density profile. Whereas lensing
probes the projected mass contained in cylinders (2D), the X-ray data is sensitive to the
spherically-averaged (3D) enclosed masses \citep[e.g.,][]{Morandi10}. The combination thus
provides information on the l.o.s.~geometry. Since the surface density of Abell 383
and the BCG isophotes are both nearly circular
($b/a \simeq 0.9$), any large departure from sphericity must be along the l.o.s. 

In detail, we adopt a triaxial gNFW form for the DM
halo:
\begin{equation}
\rho_{\DM}(r_{\epsilon,\DM}) = \frac{1}{q_\DM} \frac{\rho_0}{(r/r_s)^{\beta} (1+r/r_s)^{3-\beta}},
\label{eqn:gnfw}
\end{equation}
where
\begin{align}
r_{\epsilon,\DM}&(x,y,z) = \label{eqn:reps}\\
&\sqrt{(1-\epsilon_{\Sigma,\DM})x^2 + (1+\epsilon_{\Sigma,\DM})y^2 + (z/q_{\DM})^2}. \nonumber
\end{align}
Here the $z$-axis is the l.o.s.; the factor $1/q_{\DM}$ in Equation
\ref{eqn:gnfw} therefore ensures that the surface density is constant
as $q_{\DM}$ varies. The ellipticity of the mass surface density
$\epsilon_{\Sigma,\DM}$ is related to that of the lensing potential
$\epsilon_{\phi,\DM}$ following \citet{Golse02}. Note that $q_{\DM}>1$
and $<1$ correspond to prolate and oblate cases, respectively.

Following N09, the stellar mass of the BCG is modelled by
a dPIE profile \citep{Eliasdottir08} fit to \emph{HST} surface photometry. However, we consider a
more general triaxial deprojection, with $r_{\epsilon,*}$ defined
as in Equation \ref{eqn:reps}, replacing $\epsilon_{\Sigma,\DM}$ and
$q_{\DM}$ by $\epsilon_{\Sigma,*}$ and $q_*$, respectively.
 
Before describing our detailed analysis, it is useful to gain some physical
insight into the effects of varying $q_{\DM}$ and $q_{*}$. $q_{\DM}$ 
governs the ratio between 2D and 3D halo masses and is therefore
well-constrained by the combination of lensing and X-ray data. In
contrast, $q_{*}$ is not constrained by long-slit kinematic data, and we therefore
adopt a prior distribution based on knowledge of the intrinsic axis ratios
of elliptical galaxies \citep{Tremblay95}. We can expect
that $1 < q_* < q_{\DM}$ (in the prolate notation), both because simulated DM halos are much more
flattened ($\langle c/a \rangle \simeq 0.5$, \citealt{Jing02}) than stars in elliptical galaxies
($\langle c/a \rangle \simeq 0.7$, \citealt{Tremblay95}),
and because isotropic dissipation
processes in the baryon-dominated regime should yield rounder
mass distributions
\citep[e.g.,][]{Abadi10}. Qualitatively, we expect that for a fixed halo,
rounder stellar orbits will enclose less mass, thereby
reducing the observed stellar velocity dispersion. This
introduces a degeneracy between $q_*$ and $\beta$.
By accounting for this degeneracy, we incorporate uncertainties arising
from triaxiality and projection into our final results.

Technically, models are compared to the X-ray data by computing spherically-enclosed
masses in a triaxial mass distribution, as justified in Sections~\ref{sec:Xray} and \ref{sec:systematics}.
Fully triaxial dynamical models are not computationally feasible. However, since the observations
imply DM axis ratios of $x:y:z \simeq 1\,:1.1\,:\,2$, a spheroidal treatment with the symmetry axis along the l.o.s.~is a very good approximation.
This represents a significant improvement over our previous spherical dynamical models.
By assuming a two-integral distribution function $f(E,L_z)$,
the dynamics can be computed as described by \citet{Gavazzi05} and \citet{Qian95}.

Table 1 summarizes the model parameters and our assumed priors. 
As detailed in N09, models are proposed by Markov
Chain Monte Carlo (MCMC), and their likelihood is computed as
the product of the likelihoods of the four datasets.

\section{Results: a triaxial model with shallow inner slope}

\begin{figure*}
\centering
\includegraphics[width=0.6\linewidth]{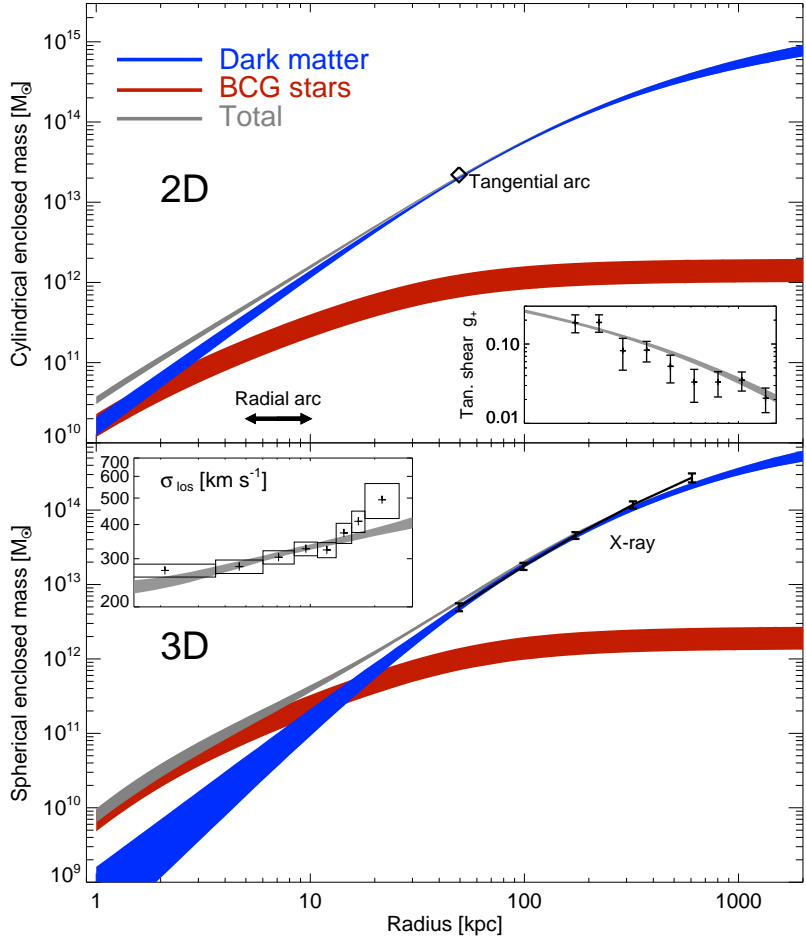}
\caption{\emph{Top:} Mass enclosed in cylinders (2D) for the dark and stellar components, as
well as the total mass distribution, with tangential reduced shear ($g_+$) data inset at the
same radial scale. \emph{Bottom:} Mass enclosed in spheres (3D), with velocity dispersion data
inset and X-ray constraints overlaid. All bands show 68\% confidence regions. The models acceptably fit all constraints from $\simeq 2$~kpc to $\simeq1.5$~Mpc.}
\label{fig:massprofile}
\end{figure*}

\begin{figure*}
\centering
\includegraphics[width=0.9\linewidth]{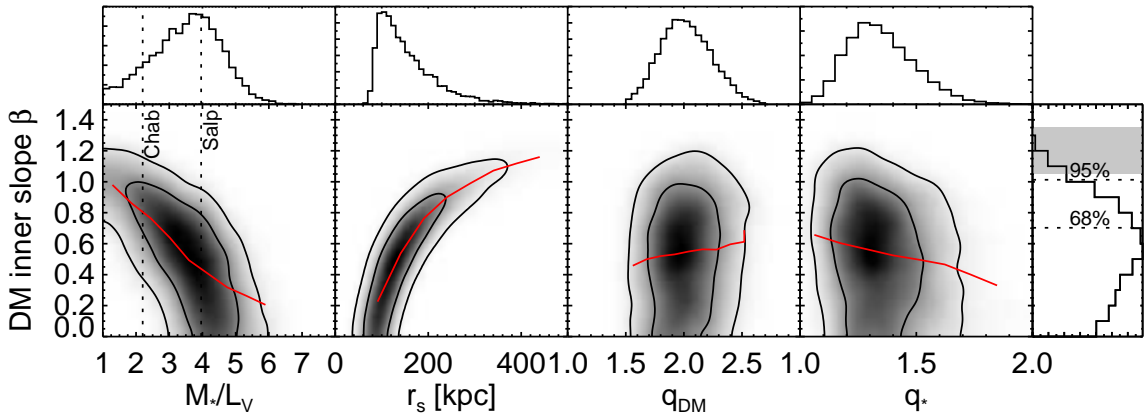}
\caption{Joint posterior probability density of $\beta$ with four other model 
parameters: $M_*/L_V$, $r_s$, $q_{\DM}$, and $q_*$. The potential degeneracies are reduced by the
inclusion of complementary constraints. Contours show $68\%$ and $95\%$ confidence regions, and red lines
indicate the mean $\beta$ to highlight the degeneracy slopes.
Histograms show marginalized posteriors. Upper limits on $\beta$ are indicated in the right panel,
along with the range (shaded) reported for cluster-scale $N$-body simulations by \citet{Diemand05}.
Dotted lines in the left panel show the $M_*/L_V$ inferred from {\it Subaru} photometry using
{\tt kcorrect} \citep{Blanton07}, assuming Chabrier and Salpeter IMFs.}
\label{fig:degen}
\end{figure*}

Our model fits are plotted in Figure~\ref{fig:massprofile} and
summarized in Table~1. To reconcile the observed
velocity dispersions with the lensing and X-ray data, shallow DM
slopes are preferred. As expected, the formal uncertainties are increased with respect
to previous models that neglected triaxiality \citep{Sand08},
yet we still obtain $\beta < 0.70$ (68\% confidence, $<1.0$ at 95\%).
Interestingly, the DM halo is
found to be elongated along the l.o.s., with $q_{\rm DM}\sim2$,
typical of simulated halos \citep{Jing02} and consistent
with the orientation selection bias expected for strong lenses.
Furthermore, the stellar
mass-to-light ratio is found to be in agreement with the values
inferred from stellar population synthesis models, assuming a Salpeter
IMF (Treu et al.\ 2010).

The parameter degeneracies are illustrated in
Figure~\ref{fig:degen}. It is instructive to see how they have been 
minimized by the combination of observational tools unique
to our method. By including only lensing constraints,
for example, we would obtain weaker constraints on
$\beta$ due to the unconstrained stellar mass.
The degeneracy with $r_s$ is reduced using weak lensing and X-ray 
probes at large radii. Finally, the DM l.o.s.~ellipticity
$q_{\DM}$ is determined by the combination of X-ray and lensing data.

Equally important to the inferred model parameters is the ``goodness
of fit.'' As Figure~\ref{fig:massprofile} clearly shows,
a relatively simple model, with a single DM halo characterized
by a simple functional form, fits all the data remarkably well, including the
extended velocity dispersion data and the detailed strong lensing
features. (The best-fitting models have image plane rms errors of $0\farcs3$.)
The velocity dispersion profile is particularly encouraging:
its shape and normalization are well-matched,
which was typically challenging using spherical
dynamical models \citep[][N09]{Sand08}. 

\section{Remaining uncertainties}
\label{sec:systematics}

In this Letter we have incorporated additional data and more sophisticated
models to address the impact of projection. Here we consider the residual systematic uncertainties.
First, we repeated the analysis with all $\sigma_{los}$ measurements shifted by
10\% to account for systematic measurement uncertainties. We note that the mild
radial orbital anisotropy typically observed at lower redshift can only decrease $\beta$ (see N09). 

Second, we assumed that the BCG is coincident with the center of
the DM halo, consistent with the $\lesssim 3$~kpc projected offset required
by the lensing. A similar 3D offset has
little effect on the enclosed mass outside $\simeq 6$~kpc, so we can evaluate
the effects of this assumption by excluding data within 6~kpc.

A spatially-constant $M_*/L$ was assumed, consistent with our non-detection
of a color gradient outside the central $\simeq 1''$.
Following \citet{Kelson02}, we estimate
limits on $\Delta M_*/L$ from those on $\Delta (B-R)$. We translate this to
an uncertainty on $R_e$ and repeat the analysis shifting $R_e$.

Finally, we recall that the A08 X-ray measurements assumed sphericity, whereas our mass models
are non-spherical. By calculating the gas emission
in a non-spherical halo with $q_{\rm DM} = 2$, we estimate that spherically deprojecting the
X-ray observables biases the inferred (spherically-averaged) mass profile by only $\simeq 7\%$ typically,
consistent with previous studies \citep{Gavazzi05,Nagai07}. As discussed
in Section~\ref{sec:Xray}, this small bias is comparable to other systematic uncertainties
inherent to X-ray analyses and is within our adopted calibration uncertainty.
We estimate the impact on our results by shifting the X-ray masses accordingly. 

In all cases, the limits on $\beta$ shifted by $<0.06$. We conclude that
the remaining known systematic uncertainties are much smaller than the projection uncertainty addressed
in this Letter.

\begin{deluxetable*}{lccc}
\tablewidth{\textwidth}
\tablecolumns{4}
\tablecaption{Models Inferred from Strong and Weak Lensing, Kinematic, and X-ray Data}
\tablehead{\colhead{Quantity} & \colhead{Units} & \colhead{Prior} & \colhead{Marginalized Posterior}}
\startdata
\multicolumn{4}{l}{\emph{gNFW DM halo}} \\
$\epsilon_{\phi,{\rm DM}}$ & \ldots       & $[0, 0.15]$   &  $0.055^{+0.017}_{-0.014}$     \\
Position angle (P.A.)                     &  deg         & $[-10, 30]$    & $10.5^{+7.6}_{-5.7}$      \\
$r_s$                    & kpc          & log-uniform*   & $112^{+61}_{-30}$      \\
$\sigma_{0,\DM}$          & km~s${}^{-1}$ & log-uniform*  & $1629^{+150}_{-125}$      \\
$\beta$                  & \ldots       & uniform* & $0.59^{+0.30}_{-0.35}$       \\
$q_{\DM}$                 & \ldots       & $[1.5, 2.7]$    & $1.97^{+0.28}_{-0.16}$      \\
\hline
\multicolumn{4}{l}{\emph{BCG stellar mass}} \\
$r_{\rm core}, r_{\rm cut}$ (dPIE) & kpc    & 0.82, 40.5 & (fixed) \\
$b/a$                           & \ldots  & 0.88     & (fixed) \\
P.A.                            & deg    & 15     & (fixed) \\
$M_*/L_{{\rm V}}$          & $(M/L_{\rm V})_{\sun}$ & [1,8] &  $3.85^{+0.90}_{-1.19}$ \\
$q_{*}$                   & \ldots        & $1/q_*^2 \sim 0.54 \pm 0.15$ & $1.30^{+0.15}_{-0.13}$ \\
\hline
\multicolumn{4}{l}{\emph{Calibration parameters}} \\
$m_{\rm WL}$               & \ldots        & $0.81 \pm 0.04$ & $0.78^{+0.03}_{-0.02}$   \\
$m_{\rm X}$                & \ldots        & $0.90 \pm 0.10$  & $1.01^{+0.04}_{-0.05}$   \\
\hline
\multicolumn{4}{l}{\emph{Cluster galaxy perturbers}} \\
$\sigma_{0,*}$             & km~s${}^{-1}$  & $159 \pm 40$      &  $122^{+18}_{-17}$    \\
$r_{{\rm cut},*}$           & kpc           & $[20,80]$          & $\dagger$    \\
$r_{{\rm cut, P1}}$         & kpc           & $[20,120]$          & $\dagger$    \\
$r_{{\rm cut, P2}}$          & kpc & $[20,80]$        & $\dagger$     
\enddata
\tablecomments{\small Posteriors are summarized using the mode and the 68\% confidence interval. These priors were found to be non-restrictive following initial tests with broader priors. $[a,b]$ denotes a uniform prior, while $\mu \pm \sigma$ denotes a Gaussian, which is truncated at $1.5\sigma$ for calibration parameters. *In practice, independent, uniform priors are placed on linear combinations of $(\log r_s, \log \sigma_{0,\DM}, \beta)$ for efficiency; these are equivalent to broad, uniform priors on $\log r_s$, $\log \sigma_{0,\DM}$, and $\beta$. The prior on $q_*$ is a fit to \citet{Tremblay95}. Formulae to convert $\sigma_0$ to other quantities are given in N09. $\sigma_{0,*}$ and $r_{{\rm cut},*}$ denote dPIE parameters for an $L_*$ cluster galaxy (see N09). Perturber radii marked with $\dagger$ have nearly flat posteriors, which are omitted.}
\end{deluxetable*}

\section{Discussion}
\label{sec:discussion}

The shallow inner DM slope we find in Abell 383 is difficult to
reconcile with results from numerical simulations. DM-only $N$-body
simulations predict $\beta \simeq 1 - 1.3$.
\citep[e.g.,][]{Diemand05}.
Although recent galaxy-scale simulations have suggested continuous, modest flattening \citep{Stadel09,Navarro10},
this is unlikely to affect our comparison, since the slope becomes shallower than NFW only on very
small scales ($\lesssim 0.015 r_s \approx 4$~kpc) that encompass only the innermost velocity dispersion bin;
this datum does not dominate our results. 

Cosmological hydrodynamical simulations incorporating baryonic
physics (cooling and feedback) with high resolution find that baryon
condensation in the cluster core steepens the DM slope (an effect
termed ``adiabatic contraction''), thereby increasing $\beta$ and 
worsening the discrepancy with observations
(e.g., \citealt{Gnedin04}, $\beta \simeq 2$; \citealt{Duffy10}, $\beta \simeq 1.5$;
\citealt{SommerLarsen09}, $\beta \simeq 1.1$). The
amount of steepening depends on the details of the subgrid implementation,
but the sign of the effect is consistent.

Interestingly, recent hydrodynamical simulations at the galaxy and
dwarf galaxy scales have shown that baryons can {\it soften} DM cusps
\citep[e.g.,][]{Governato09,RomanoDiaz09}; however, the relevant processes
appear not to scale to galaxy clusters, which have deeper potential
wells than dwarf galaxies and are less baryon-dominated than $L_*$
galaxies. Processes that have been suggested to counteract adiabatic
contraction in cluster cores, such as a late dry merging
\citep{Gao04} or dynamical friction by infalling baryonic clumps
\citep{ElZant01,ElZant04,Nipoti04}, are apparently subdominant in cosmological
hydrodynamical simulations and do
not lead to a shallow cusp with $\beta < 1$. Our results on shallow DM cusps in galaxy
clusters thus seem to require a revision of our understanding of either the DM backbone or
the most relevant baryonic physics for shaping the cluster core.

\section{Summary}
\label{sec:conclusions}

We have extended our previous analyses of Abell 383 by incorporating weak lensing and 
X-ray data. Based on deep Keck spectroscopic observations, we
have also measured -- for the first time in a lensing cluster -- an
extended velocity dispersion profile that demonstrates a clear rise 
in the outer regions in response to the cluster potential. As a result of these improved
datasets, we have refined our modelling to consider triaxial mass distributions.
We demonstrate that all four observational ingredients are
essential to obtain a complete three-dimensional view of Abell 383
over a very wide dynamic range of three decades in radius.

We find that the DM distribution in Abell 383 is clearly triaxial, consistent 
with $\Lambda$CDM numerical simulations. However, its DM profile has a shallow
density cusp with $\beta < 1$ (95\%), a result that appears
inconsistent with numerical simulations of clusters in a $\Lambda$CDM context at moderate significance.
Although Abell 383 represents only a single well-studied cluster,
comparable data are currently in hand to extend this analysis to sample of
9 clusters whose ensemble properties will be presented in a forthcoming paper.

\acknowledgments
It is a pleasure thank Steve Allen for providing his X-ray analysis and to
acknowledge the helpful assistance of Eric Jullo, Johan Richard, Jean-Paul Kneib,
and Satoshi Miyazaki. We thank the referee, Marceau Limousin, for his constructive suggestions.
R.S.E.~acknowledges financial support
from DOE grant DE-SC0001101.  Research support by the Packard
Foundation is gratefully acknowledged by T.T.  The authors wish to
recognize and acknowledge the cultural role and reverence that the
summit of Mauna Kea has always had within the indigenous Hawaiian
community.  We are most fortunate to have the opportunity to conduct
observations from this mountain.

\bibliographystyle{apj}

\end{document}